# A Model-Driven Digital Twin for the Systematic Improvement of DevOps Pipelines

Achref Samoud
Sara Aissat
achref.samoud.1@ens.etsmtl.ca
sara.aissat.1@ens.etsmtl.ca
ÉTS | École de technologie supérieure
Montreal, Canada

Francis Bordeleau
francis.bordeleau@etsmtl.ca
ÉTS | École de technologie supérieure
Montreal, Canada

Ali Tizghadam
ali.tizghadam@telus.com
TELUS
Toronto, Canada

**Abstract**

CI/CD pipelines are central to DevOps practices, yet their growing complexity makes them increasingly difficult to interpret, analyze, and systematically evolve. Existing tooling primarily offers execution logs and static graph representations, providing limited support for structured analysis of pipeline behavior, failures, and version-to-version evolution. This paper presents a model-driven Digital Twin (DT) for CI/CD pipelines that leverages BPMN as a modeling backbone to transform raw CI configurations into structured, higher-level process representations. The proposed DT architecture enables visual abstraction of pipeline structure, failure tracing, and systematic version comparison, supporting both monitoring and evolution analysis of DevOps processes. Building upon validated DT architectural principles and prior work on build optimization and anomaly detection, the framework provides a modular, extensible foundation for integrating advanced analytical and prescriptive services into software delivery processes. The approach is validated using open-source CI/CD projects, and ongoing work targets the integration of additional improvement services and the extension of the DT to broader DevOps lifecycle processes.

## 1 Introduction

DevOps has emerged as a software engineering paradigm aiming to bridge development and operations in order to enable rapid, reliable, and continuous software delivery [9, 17]. It promotes automation, cross-functional collaboration, and continuous feedback across the lifecycle of software systems, with the goal of reducing deployment friction and improving delivery performance [18] through automation at every stage, from planning to release, including development, testing, integration, release preparation (building/packaging), and monitoring [23]. CI/CD pipelines play a central role in the DevOps approach to automate the execution of the software process phases to reduce the time to deliver new features and increase software quality.

Empirical studies of CI/CD usage in large-scale open-source projects show that pipeline configurations undergo frequent evolution throughout a project's lifetime, driven by dependency changes, build matrix adjustments, and environment reconfiguration [22]. However, this evolution is typically conducted in an ad hoc manner, relying on intuition rather than empirical data. Consequently, this leads to a significant waste of resources, effort, and costs [5]. Furthermore, the lack of dedicated tools to support systematic pipeline evolution makes it difficult to measure the impact of modifications, frequently resulting in issues that are hard to detect, diagnose, and resolve.

DTs have emerged over the last decades as a paradigm for creating virtual representations of physical or operational systems that are continuously synchronized with their real-world counterparts [20, 31]. When applied to processes rather than physical assets, DTs fundamentally rely on models to represent the structure, behavior, and evolution of the system being twinned, making them inherently model-driven constructs. In this work, we leverage model-driven engineering (MDE) and BPMN [25] as the modeling backbone for developing a DT that supports systematic pipeline improvement.

Our research, conducted within the context of an industrial research chair in DevOps, aims to develop a Digital Twin (DT) for the systematic evolution of various DevOps processes.

In this paper, we focus specifically on a DT tailored for CI/CD pipelines. The remainder of the paper is structured around its three main contributions. First, we detail the MDE foundations of the DevOps pipeline DT, including its underlying metamodel and model transformation that maps CI/CD configurations to BPMN 2.0 process models (Section 3). Second, we describe the DT architecture and the implementation of its foundational components and services, including ongoing work to develop and integrate various pipeline improvement services (Section 4). Third, we present the validation of the DT conducted using open-source projects and outline its current limitations (Section 5). Finally, we discuss related work (Section 6) and outline our future research plans to expand the DT to other facets of the DevOps lifecycle (Section 7).

## 2 Background

### 2.1 DevOps pipelines as software processes

Central to DevOps is the notion of a *CI/CD pipeline*, an automated workflow that encompasses Continuous Integration (CI), and Continuous Delivery (CD). CI focuses on the early phases of the delivery cycle: developers frequently integrate code changes into a shared repository, triggering automated build and validation workflows. Each integration is verified by compiling the code, running static analysis, executing unit and integration tests, and producing build artifacts. Empirical studies show that CI adoption is associated with improved defect detection, faster release cycles, and higher software quality [22, 29, 32]. CD extends CI by automating the preparation of release-ready artifacts packaging, environment provisioning, staging deployment, and acceptance testing, so that any successfully validated commit could be deployed to production and released at the push of a button.

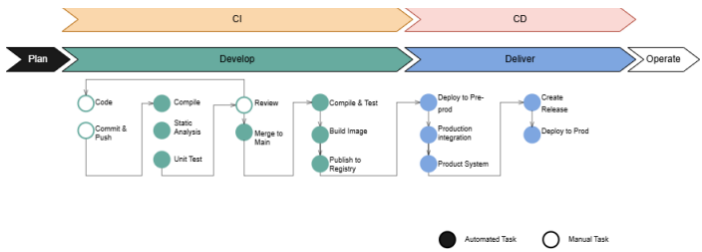

**Figure 1: DevOps Process.**

Figure 1 provides an overview of the DevOps process, showing how CI and CD activities are organized across the development and delivery phases and how automated and manual tasks alternate throughout the workflow.

In practice, this process is encoded as a declarative configuration file (e.g., `.gitlab-ci.yml`, GitHub Actions YAML, `Jenkinsfile`) that specifies an ordered sequence of *stages*, each containing one or more *jobs*. Figure 2 illustrates this with a representative GitLab CI configuration. A typical CI/CD pipeline progresses through build, test, package, deploy-staging, and deploy-production stages, with each stage orchestrating interdependent activities managed by the CI/CD engine and executed on distributed runners.

## 2.2 BPMN for structured process modeling

Business Process Model and Notation (BPMN) is a graphical standard maintained by the Object Management Group (OMG) for specifying business processes [25]. Originally developed to bridge the communication gap between business analysts and technical implementers, BPMN has become the de facto notation for process modeling in both industry and academia.

BPMN supports various types of analysis by providing a visual foundation for understanding and improving business processes. Through process flow analysis, it helps identify bottlenecks, redundancies, and inefficiencies in the sequence of activities, while gap analysis allows organizations to compare the current state (As-Is) process with the desired future state (To-Be) to pinpoint areas for improvement. It also enables root cause analysis by tracing process failures back to their origin, and compliance analysis to verify whether processes adhere to regulatory requirements and standards. When combined with performance metrics, BPMN supports performance analysis to measure efficiency in terms of time, cost, and resources, as well as risk analysis to identify potential failure points at decision gateways. Additionally, many BPMN tools support simulation analysis, allowing organizations to test and validate process changes before actual implementation.

Beyond its origins in business process management, BPMN has been adopted in diverse technical domains. [14] provide a comprehensive overview of the standard and its adoption across sectors, while [26] extend BPMN specifically for software process tailoring, demonstrating its applicability to software engineering workflows. These studies confirm that BPMN's expressiveness generalizes well

```yaml
 1  default:
 2    image: maven:3.9-eclipse-temurin-21
 3    services: [docker:24-dind]
 4  variables:
 5    REGISTRY: $CI_REGISTRY_IMAGE
 6    DEPLOY_ENV: staging
 7  workflow:
 8    rules:
 9      - if: $CI_PIPELINE_SOURCE == "push"
10      - if: $CI_PIPELINE_SOURCE == "merge_request_event"
11  stages: [build, test, package, deploy]
12  .ci_template:
13    retry: 2
14    tags: [docker]
15  compile:
16    extends: .ci_template
17    stage: build
18    script: mvn compile
19  static-analysis:
20    extends: .ci_template
21    stage: build
22    script: sonar-scanner
23    rules:
24      - changes: [src/**, pom.xml]
25  unit-test:
26    stage: test
27    script: mvn test
28  build-image:
29    stage: package
30    image: docker:24
31    script:
32      - docker build -t $REGISTRY .
33      - docker push $REGISTRY
34    needs: [compile, unit-test]
35  deploy:
36    stage: deploy
37    script: kubectl apply -f k8s/$DEPLOY_ENV/
38    when: manual
39    environment: $DEPLOY_ENV
40
```

**Figure 2: Representative .gitlab-ci.yml pipeline configuration**

to domains where process structure must be explicitly represented and analyzed.

## 2.3 DTs for software processes and DevOps processes

The concept of DT originated in manufacturing, where [19] introduced it as a virtual replica of a physical system that mirrors its state, behavior, and lifecycle. A DT comprises three core components: a physical twin, a virtual counterpart, and a bidirectional data connection enabling synchronization between the two. [8] provide a comprehensive survey of DT definitions, characteristics, and design implications, identifying three foundational capabilities that distinguish a DT from a conventional simulation: (1) *continuous synchronization* between the physical and virtual entities, (2) *monitoring and analysis* of the physical system's state through its virtual counterpart, and (3) *predictive and prescriptive* reasoning enabled by the living model. The paradigm has expanded well beyond manufacturing into essentially all application domains, including built assets, healthcare, smart cities, infrastructure management, and, more recently, software systems.

Within software engineering, the DT concept has gained traction as a means to represent and reason about complex, evolving systems. [21] present a systematic literature review of DTs applied to software systems, identifying key themes: runtime monitoring, architecture-level simulation, and model-driven synchronization between deployed systems and their virtual representations. Their analysis highlights that a software-oriented DT typically relies on a *model layer* that abstracts the system's structure, a *data layer* that captures runtime behavior, and a *service layer* that exposes analytical capabilities over both.

Applying these DT principles to DevOps processes, and to CI/CD pipelines in particular, requires identifying what constitutes the physical entity, the virtual counterpart, and the synchronization mechanism in this context. It also requires instantiating the three architectural layers (model, data, service) for the specific characteristics of pipeline definitions and their executions. These dimensions provide the conceptual foundation upon which the present work builds its DT for DevOps pipelines.

## 3 Model-driven DevOps pipeline

This section presents an MDE approach for representing CI/CD pipelines as structured process models. We first motivate the need for structural analysis beyond execution logs, then formalize the pipeline metamodel, detail the transformation from GitLab CI configurations to BPMN, and describe the analytical services enabled by this representation.

### 3.1 Motivation

Existing CI/CD platforms such as GitLab CI, GitHub Actions, and Jenkins provide execution dashboards that display job status, logs, and dependency graphs. While useful for immediate troubleshooting, these views are tightly coupled to individual pipeline executions and offer limited support for reasoning about *pipeline structure* as a first-class artifact. Specifically, three analytical needs remain unaddressed:

(1) **Structural comprehension:** As pipelines grow in complexity, developers struggle to understand the overall workflow topology from configuration files alone. A declarative YAML definition does not explicitly convey flow semantics.
(2) **Version comparison:** CI/CD configurations evolve alongside the codebase, yet existing tools offer no structured mechanism to compare *pipeline structures* across versions. Textual diffs of YAML files do not capture semantic changes, making it difficult to assess the impact of configuration changes on pipeline behavior.
(3) **Evolution tracking and causal reasoning:** CI/CD pipelines evolve over time, and their execution outputs vary across versions. Current CI/CD platforms provide per-run metrics, but offer no structured mechanism to correlate performance regressions or recurring failures with specific structural changes in the pipeline definition.

These limitations motivate the adoption of a model-driven approach [12], in which CI/CD configurations are systematically transformed into a standardized process modeling notation capable of supporting visual abstraction, structural comparison, and execution overlay. In line with model-driven engineering principles, the goal is to elevate CI/CD pipeline definitions from opaque configuration artifacts to explicit, analyzable and actionable process models. We adopt BPMN 2.0 [25] as the target notation, based on its semantic alignment with CI/CD pipeline constructs, as detailed below.

The analytical goals outlined above for structural comprehension, version comparison, and causal reasoning require more than ad hoc parsing of YAML files. They demand a formal representation of CI/CD pipelines that is stable enough to support cross-version comparison and rich enough to capture both structure and runtime behavior. This subsection presents the metamodel that serves this role. Figure 3 provides an overview of the complete metamodel; the remainder of this section discusses the reasoning behind its structure.

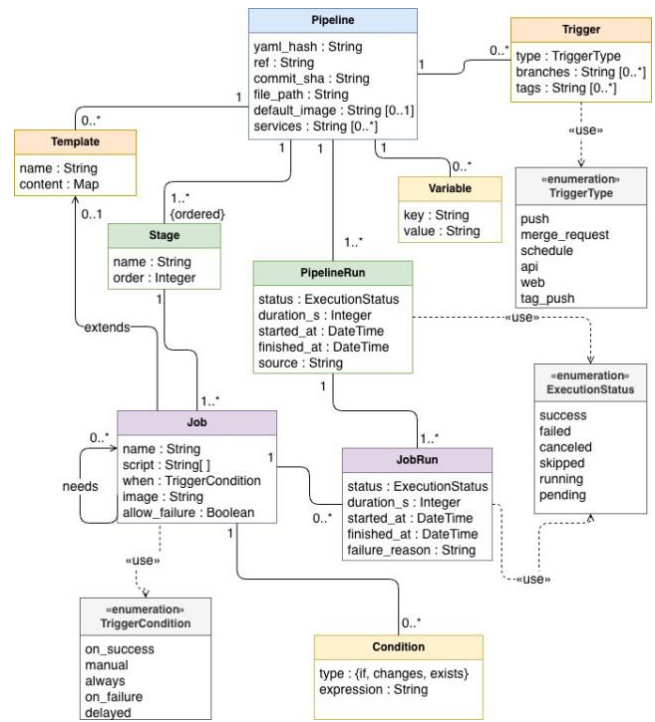

**Figure 3: Pipeline metamodel**

### 3.2 Pipeline metamodel

The analytical goals outlined above, demand a formal representation of CI/CD pipelines . This subsection presents the metamodel that serves this role.

**From configuration artifact to structural model:** Every CI/CD pipeline originates as a declarative configuration artifact a `.gitlab-ci.yml` committed to a version-controlled repository. The metamodel is accordingly organized in a *Pipeline* entity that encapsulates the provenance of this artifact: the branch reference (*ref*), the originating commit (*commit_sha*), the repository-relative file path (*file_path*), and most importantly a content-based hash of the YAML body (*yaml_hash*). This last attribute constitutes a foundational design decision: by deriving pipeline identity from content rather than from execution identifiers, the metamodel decouples structural elements from runtime instantiation. Consequently, two distinct commits that yield identical YAML content are recognized as representing the same structural version, a property that underpins meaningful cross-version comparison.

The primary structuring element within a pipeline definition is the *Stage*. In the default model, stages impose synchronization barriers: all jobs belonging to stage $S_i$ must finish before any job in stage $S_{i+1}$ can start. A Pipeline is therefore composed of an ordered sequence of stages, and this total ordering defines the macro-level control flow of the workflow in the absence of explicit inter-job dependencies. Each *Stage*, aggregates one or more *Jobs*, which constitute the atomic units of computation. A *Job* encapsulates a script to execute, an execution environment specification (*image*), a *when* trigger condition governing its activation semantics (`on_success`, `manual`, `always`, `on_failure`, `delayed`), and an *allow_failure* flag that determines whether a job failure propagates to the pipeline level.

Within a given stage, jobs that share no explicit ordering constraints are considered independent and execute concurrently reflecting GitLab CI's default scheduling semantics. However, pipelines also employ fine-grained ordering through GitLab's Directed Acyclic Graph (DAG) mode: the `needs:` directive permits a job to declare explicit precedence constraints on other jobs, irrespective of stage boundaries. A central modeling decision arises here: rather than reifying these inter-job dependencies as a distinct entity, which would introduce an intermediate class carrying no intrinsic attributes, the metamodel represents them as a *needs* self-association on *Job*.

Two supplementary structural dimensions complete the definition side. First, pipelines may declare *Variables*, key-value pairs that parameterize execution behavior (e.g., registry URLs, feature flags, environment selectors). The Pipeline entity owns a set of *Variables*. Second, each *Job* may be governed by zero or more *Conditions*: rule-based predicates (`if`, `changes`, `exists`) that the CI engine evaluates at pipeline instantiation time to determine whether a job participates in a given execution. A deliberate modeling choice dictates that *Conditions* are represented as job-level metadata rather than as structural control-flow elements; the BPMN representation consequently depicts all *defined* jobs, while execution-time filtering is conveyed through status overlays.

The metamodel further promotes pipeline *Triggers* to first-class entities. A pipeline is not merely "executed", it is *initiated* by a specific event source: a code push, a merge-request creation, a scheduled invocation, an API call, or a manual activation. CI/CD platforms encapsulate this information within `workflow:rules` expressions or legacy `only/except` directives, rendering it opaque in any structural representation. By extracting triggers as explicit entities, each typed by a *TriggerType* enumeration (`push`, `merge request`, `schedule`, `api`, `web`, `tag_push`).

**Bridging definition and execution:** The definition side of the metamodel addresses the question of what a pipeline *prescribes*. However, the analytical objectives motivating this work (failure diagnosis, performance regression detection, etc.) necessitate answering a complementary question: what did the pipeline *produce* at runtime? A metamodel confined to static structure would leave execution telemetry isolated within CI/CD dashboards, disconnected from the structural model that contextualizes it.

The metamodel therefore introduces the concept of a pipeline execution, called *PipelineRun*, facet that mirrors the definition-side hierarchy. Each instantiation of a pipeline produces a *PipelineRun*, which captures what happened at runtime: the overall outcome, how long the pipeline took, when it started and finished, and what event triggered it. The outcome is classified by an *ExecutionStatus* enumeration whose values reflect the states reported by the CI engine. A *Pipeline* is associated with one or more *PipelineRuns*.

Each *PipelineRun* is in turn composed of one or more *JobRuns*, each corresponding to a single execution instance of a *Job* within that pipeline run. A *JobRun* captures the same temporal and status attributes as its parent *PipelineRun*, supplemented by an optional *failure_reason* that preserves the CI engine's diagnostic classification.

The definition and execution facets constitute the structural foundation of the DT's analytical capabilities: they enable execution data to be projected onto any structural version of a pipeline, facilitate performance comparison across multiple runs of an identical definition, and support causal reasoning between structural modifications and behavioral regressions.

### 3.3 From declarative syntax to visual process semantics

The metamodel defined in 3.2 provides a stable vocabulary for reasoning about CI/CD pipeline structure and behavior, but a vocabulary alone does not produce analyzable diagrams. Turning a raw `.gitlab-ci.yml` into an interactive BPMN representation requires a systematic, deterministic transformation that preserves the structural semantics of the source configuration while expressing them in the graphical vocabulary of BPMN 2.0. This section presents the semantic correspondence between the two formalisms, describes the algorithm that carries out the transformation, and discusses the design trade-offs that shape the resulting process model.

*3.3.1 Semantic correspondence.* Table 1 presents the element-level correspondence between GitLab CI constructs and their BPMN 2.0 counterparts. The mapping goes beyond syntax: each rule reflects the alignment between the source construct's role in the CI workflow and the target element's semantics in the BPMN specification.

*3.3.2 BPMN process model generation.* Given a CI/CD pipeline YAML configuration, the transformation produces a complete BPMN 2.0 XML document which include both the process model and its diagram layout. The algorithm first separates job definitions from reserved directives and reusable templates, merges the declared stage ordering with per-job stage annotations, and extracts typed pipeline triggers from the workflow rules. A directed dependency graph is then built from `needs:` declarations, and jobs are topologically sorted within each stage to determine their placement in a column-based layout. Each job is mapped to the appropriate BPMN activity type (`userTask` for manual jobs, `task` for all others) and assigned to a lane representing its stage. Stages containing two or more independent jobs receive a `fork-join` parallel gateway pair that captures intra-stage concurrency. `Start` events are generated according to the extracted triggers: a single untyped event when no triggers are present, or one typed event per trigger (signal for push, schedule, and tag; message for merge-request) merged through an exclusive gateway. `End` events are attached to terminal jobs. Inter-job dependencies declared through `needs` are rendered using parallel gateways: a join gateway collects multiple incoming edges before a job with several predecessors, while a split gateway distributes control after a job that fans out to multiple same-stage successors. When all predecessors reside in earlier stages, the transformation reuses the existing stage-level join gateway rather than introducing additional gateway nodes, thereby reducing visual clutter. Single dependencies produce direct sequence flows. Consecutive stages are linked by sequence flows whose endpoints adapt to the presence or absence of parallel gateways on either side. Finally, a deterministic layout algorithm assigns each stage to a fixed-width column, vertically stacks jobs within dynamic-height lanes, and routes edges along vertical bus lines flanking each column. Figure 4 shows the BPMN process model produced by applying this transformation to the pipeline configuration of Figure 2.

Table 1: Semantic mapping from GitLab CI constructs to BPMN 2.0 elements.

| GitLab CI construct | BPMN 2.0 element | Semantic rationale |
| --- | --- | --- |
| Pipeline | bpmn:process | Entire CI workflow maps to a single executable BPMN process. |
| Stage | bpmn:lane | Lanes partition the diagram by execution phase, grouping co-stage jobs. |
| Automatic job | bpmn:task | Generic automated activity; applies when when ≠ manual. |
| Manual job | bpmn:userTask | Human-triggered activity, modeling manual approval gates. |
| Intra-stage parallelism | bpmn:parallelGateway | AND-split/join pair capturing concurrent execution within a stage. |
| needs: dependency | bpmn:sequenceFlow | Explicit ordering constraint preserved for inspection but excluded from rendering. |
| Stage ordering | bpmn:sequenceFlow | Sequential barrier between consecutive stages. |
| Pipeline trigger | Typed bpmn:startEvent | signalEventDefinition for push/schedule/tag; messageEventDefinition for merge-request. |
| Pipeline end | bpmn:endEvent | Standard BPMN delimiter marking process termination. |
| Job script | bpmn:documentation | Script content embedded as documentation within the task element. |

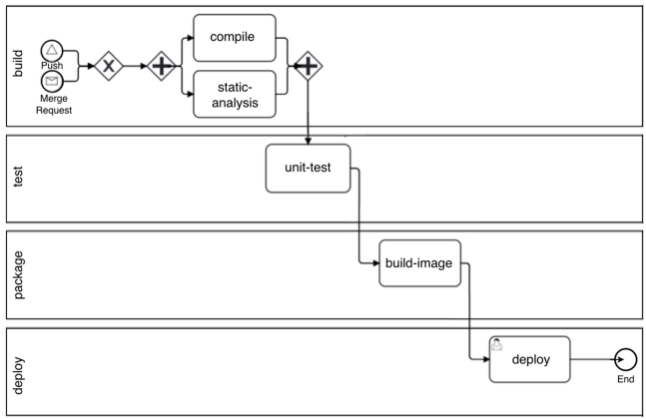

**Figure 4: BPMN process model generated from the CI configuration in Figure 2.**

## 4 DevOps pipeline DT

The MDE approach presented in the previous section established the foundations for representing CI/CD pipelines through a metamodel, a transformation to BPMN, and a set of associated analytical capabilities. While this representation enables structural abstraction and analysis, it remains insufficient on its own to support continuous interaction with evolving pipeline instances. To make these models actionable, they must be embedded within a system capable of continuously acquiring pipeline data, organizing both structural and runtime information, and exposing services for analysis and interaction. This role is fulfilled by the DevOps pipeline DT, which operationalizes these foundations within a unified and extensible system. The implementation of this DT is publicly available as open source[1].

### 4.1 Objectives and overview of the DT

The primary objective of the DevOps pipeline DT is to provide a living representation of CI/CD pipelines that integrates both structural and runtime knowledge within a coherent and extensible system. To achieve this objective, the DT is designed as a set of coordinated components and services that collectively support the acquisition, management, and exploitation of pipeline-related information.

More specifically, the DT is intended to support four complementary objectives: (1) enabling the acquisition of pipeline definitions and execution data from GitLab through dedicated ingestion mechanisms, (2) maintaining coherent and accessible structural and runtime representations of pipelines over time, (3) supporting visualization, failure interpretation, and version comparison services, and (4) providing an extensible foundation for the integration of advanced analytical capabilities and eventually other DevOps process dimensions.

Figure 5, adapted from [6], presents an overview of the DT and its main functional components, which are detailed in the following subsections.

The DT is structured around the interaction between the *Actual Twin* (AT) and a set of coordinated components responsible for data acquisition, management, distribution, and service execution. The *Actual Twin* corresponds to the external DevOps environment, from which pipeline definitions and execution data are continuously retrieved. These data are ingested through dedicated acquisition components within the *DT Data Management* layer, which is responsible for collecting, structuring, and persisting both raw and processed information.

At the core of the system, an internal data distribution mechanism based on a publish/subscribe paradigm enables communication between components across the DT. This mechanism ensures loose coupling and preserves the autonomy of individual services while ensuring consistent data exchange throughout the system.

---
[1] https://gitlab.com/devops-pipeline-dt

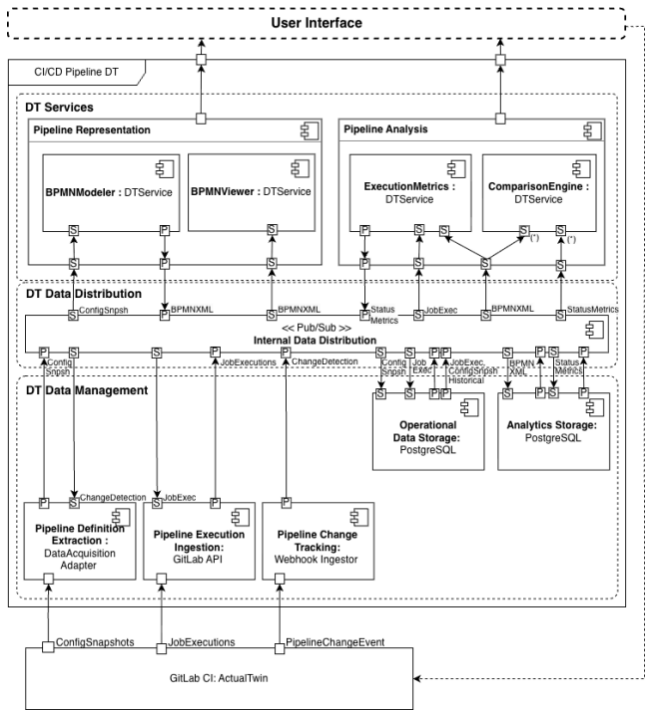

**Figure 5: Architecture of the DevOps Pipeline DT system**

On top of this foundation, the *DT Services* layer is organized into complementary service aspects aligned with the objectives defined previously. These services operate on the managed data to provide functionalities such as model generation, visualization, comparison, and analytical processing.

The following subsections detail these elements, starting with data acquisition and event handling.

### 4.2 Data Acquisition and Event Handling

Once the objectives of the DT are established, data acquisition constitutes the entry point for enabling the corresponding services.

As illustrated in Figure 5, the DT relies on a set of ingestion components in the *DT Data Management* layer responsible for retrieving pipeline-related information from the the AT, which, in this context, corresponds to the GitLab CI environment.

The acquisition process is initiated when a project is selected through the *User Interface*. The system then retrieves pipeline definitions, including YAML configuration files, associated metadata, and version information through the *Pipeline Definition Extraction* component which relies on the GitLab API to retrieve the corresponding data from the *Actual Twin*.

At this stage, the modeling logic is triggered to transform a given pipeline definition into its corresponding BPMN representation. This transformation is associated with the *configSnapshot* flow shown in Figure 5, which captures structural versions of pipelines over time. Depending on the request, this historical acquisition can target a limited number of recent versions or extend to the full set of available configurations.

These definitions are propagated through the system via an event-driven communication mechanism based on a publish/subscribe (*Pub/Sub*) infrastructure, ensuring decoupled and asynchronous data exchange between components. Through this mechanism, the acquired data are both routed to the services that require them and to the corresponding storage adapters responsible for persisting them for subsequent use.

The retrieved pipeline definitions are not directly consumed as raw configuration artifacts. They are previously interpreted and structured according to the metamodel introduced in section 3, which provides a unified representation of pipeline structure and execution concepts. This intermediate representation serves as a common foundation across the DT, ensuring consistency in how pipeline-related information is organized, stored, and exploited by services.

In this sense, the metamodel acts as a structuring layer that decouples data acquisition from its subsequent usage. While the present work leverages this representation to generate BPMN models, it is important to note that alternative representations could be derived from the same underlying model. This design choice ensures flexibility in how pipelines are interpreted and enables the integration of additional services without modifying the acquisition process.

In addition to the acquisition of pipeline definitions, a second acquisition phase is triggered when execution-related information is required to support state-based analysis and visualization. In particular, once a structural representation has been generated, additional data related to job executions (e.g., status, duration, failure information) may be retrieved from the AT through the *Pipeline Execution Ingestion* component shown in Figure 5. These data enable the DT to reflect not only the structure of pipelines but also their runtime behavior. The same event-driven mechanism is used to propagate execution data to the relevant services.

Overall, this acquisition process ensures continuous synchronization between the DT and the evolving state of CI/CD pipelines, enabling subsequent modeling, visualization, and analytical services. In particular, the *Pipeline Change Tracking* component enables continuous monitoring of the current pipeline version, allowing detected changes to trigger a renewed acquisition process (through the publication of a *ChangeDetection* event via the *DT Data Distribution* layer) and keep the DT aligned with the latest state of the AT.

### 4.3 Model management and storage

To support analysis and interaction over time, the DT maintains both structural and execution-related data in a persistent and organized manner. A first category of storage is handled through the *Operational Data Storage* adapter, which are responsible for persisting pipeline definitions and execution data once they have been retrieved from the AT and structured according to the metamodel. This storage enables efficient reuse of previously acquired information, allowing subsequent requests to directly access the required data without triggering redundant acquisition processes. In addition, an *Analytical Storage* adapter is used to persist derived artifacts produced by DT services. In particular, this includes

BPMN representations (propagated through the *BPMNXML* publication/subscription flow) generated by the *BPMNModeler*, which can be retrieved to support visualization of a given pipeline version or to enable comparison across multiple versions. By storing these intermediate results, the DT avoids repeated transformations and supports more efficient analytical operations.

Together, these storage mechanisms ensure consistent access to both operational representations and derived analytical artifacts, enabling traceability, efficient querying, and evolution analysis across pipeline versions.

### 4.4 Services and interaction

The DT provides a set of services that operate on the managed structural and execution-related information to support visualization, interaction, and analysis of CI/CD pipelines. These services are organized into complementary aspects that reflect different levels of analytical capabilities.

A first aspect, referred to as *Pipeline Representation*, focuses on the structural representation and visualization of pipelines. As illustrated in Figure 5, this aspect includes the *BPMN Modeler*, which generates BPMN representations from the metamodel-based pipeline descriptions, and the *BPMN Viewer*, which enables users to interactively explore these representations. The viewer is implemented using the bpmn-js library [13] from the bpmn.io toolkit[2], an open-source JavaScript library for rendering and interacting with BPMN 2.0 diagrams. This aspect provides a consistent and interpretable view of pipeline structures across versions.

A second aspect, referred to as *Pipeline Analysis*, builds on these representations to support reasoning about pipeline behavior. It includes services such as *ExecutionMetrics*, which overlays execution data onto the structural representation to reflect the status of jobs within a given pipeline version, and *ComparisonEngine*, which enables the identification of structural and behavioral differences across pipeline versions.

In particular, the *ComparisonEngine* supports structural comparison across pipeline versions through three phases: *extraction*, *differencing*, and *visual projection*.

**Phase 1: Model extraction.** Each pipeline definition is parsed from its YAML form into a metamodel instance capturing the entities introduced in Section 3.2.

**Phase 2: Structural differencing.** Given two metamodel instances, $V_1$ and $V_2$, the engine identifies added and removed entities through set differences, and detects modified shared entities through field-level comparison. For jobs, this includes key attributes such as stage, image, dependencies, trigger conditions, execution rules, scripts, and associated variables. The comparison also produces quantitative summaries, such as stage and job count deltas.

**Phase 3: Visual projection onto BPMN.** The detected changes are projected onto the BPMN models generated as described in Section 3.3. Added, removed, and modified jobs are highlighted through color-coded overlays, and the two versions are displayed side by side to make structural evolution explicit. This projection connects textual configurations, metamodel-level differencing, and process-level visualization.

---
[2]https://bpmn.io

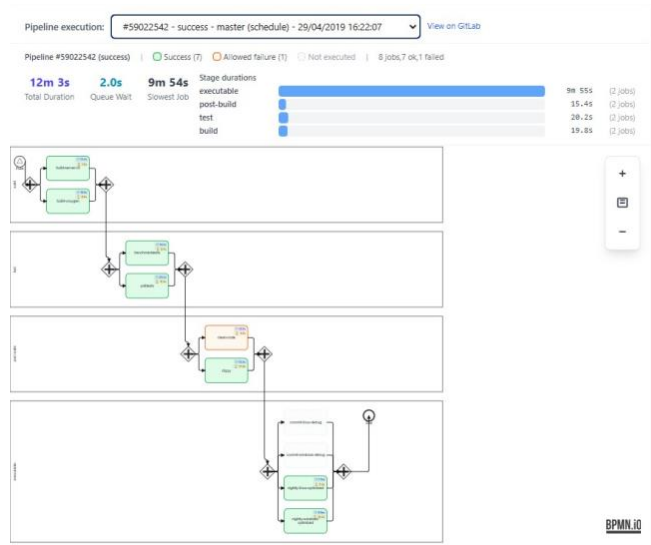

**Figure 6: BPMN-based visualization of a CI/CD pipeline with execution metrics overlay**

Together with pipeline representation and execution-aware analysis, this comparison capability constitutes one of the main service aspects currently supported by the DT. Its use is enabled through the user interface described next.

**User Interface (UI)**

The DT services are exposed through a user interface that enables interactive exploration and analysis of CI/CD pipelines. Through this interface, users can select projects, navigate across pipeline versions, and trigger visualization and comparison services.

The UI integrates the *BPMN Viewer* to provide an interactive representation of pipeline structures. Figure 6 illustrates an example of this BPMN-based visualization enriched with execution-related information. For a selected pipeline version and one of its executions, this representation can be enriched through the *ExecutionMetrics* logic, which overlays runtime metrics such as job status, duration, and failure indicators onto the BPMN model. This execution overlay allows users to visually correlate structural elements to the behavior observed during a specific pipeline run.

In addition to the process-level visualization, the interface provides complementary dashboard views that present aggregated metrics and performance indicators across multiple executions of a given pipeline version. These views support higher-level analysis by summarizing success and failure rates, duration statistics, stage-level performance, and recurring failure patterns, thereby facilitating both inspection and informed decision-making. Figure 7 illustrates an example of such a dashboard view.

Beyond execution-oriented information, the interface also supports structural exploration of pipelines through their BPMN-based representation. This allows users not only to inspect the structure of an individual pipeline version, but also to compare two versions in order to identify structural differences over time. This functionality is supported by the *ComparisonEngine* service, which highlights added, removed, or modified elements across BPMN models according to the comparison logic introduced earlier, thereby making pipeline evolution more explicit and interpretable. Figure 8 illustrates an example of such structural comparison between two pipeline versions.

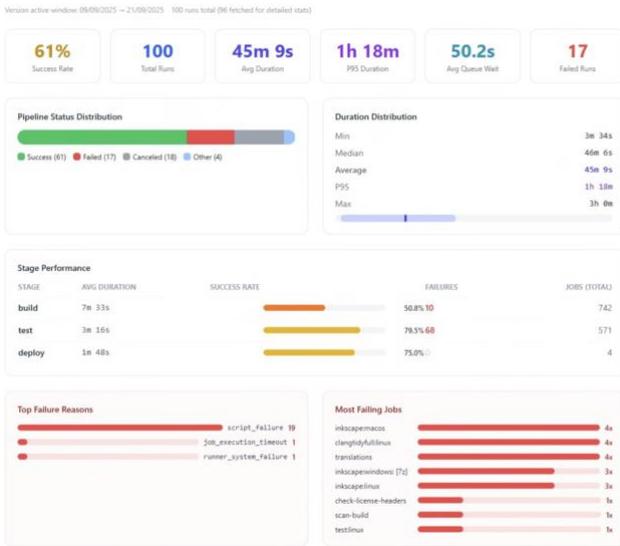

Figure 7: Dashboard view presenting aggregated pipeline metrics and execution statistics

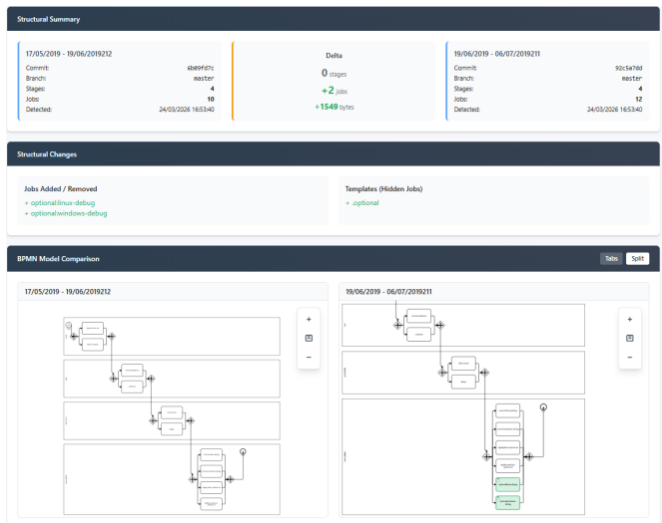

Figure 8: Structural version comparison between two pipeline versions

At this stage, any action applied back to the *Actual Twin* remains human-in-the-loop, as represented by the dashed arrow from the *User Interface* to the AT in Figure 5, with the user serving as the final decision point before any change is enacted. This interaction mode reflects the current scope of the DT, which focuses on observation, interpretation, and comparison across both structural and execution-related views of CI/CD pipeline versions, while also providing a continuously updated view of the latest pipeline version.

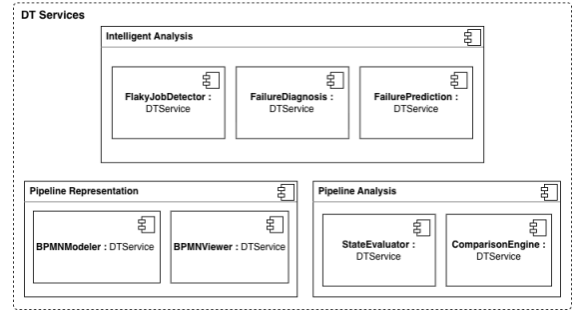

Figure 9: DT services organized by analytical aspects

**Extensibility of the DT**

Beyond descriptive and analytical capabilities, the proposed DT is designed to support the integration of more advanced, learning-based services. As illustrated in Figure 9, an additional aspect, referred to as *Intelligent Analysis*, extends the DT with capabilities derived from recent work on pipeline failure analysis [3].

This aspect includes a *FlakyJobDetector*, which identifies intermittent job failures using few-shot learning [1], a *FailureDiagnosis* service, which assigns failure instances to meaningful categories to support diagnosis and prioritization [2], and a *FailurePrediction* service, which anticipates failure categories to enable proactive decision-making in pipeline management [4].

## 5 Validation

This section assesses the proposed model-driven DT from a modeling perspective, focusing on its ability to capture, represent, and support reasoning about CI/CD pipeline structure and evolution.

### 5.1 Case Study Context

The approach was applied to three open-source GitLab projects: Wireshark[3], Veloren[4], and Inkscape[5], selected according to three criteria: (i) the complexity of their CI/CD configurations, which collectively exercise a broad range of GitLab CI constructs (stages, DAG dependencies, conditional rules, triggers, templates), (ii) their sustained development activity, which provides a rich history of pipeline evolution across commits, and (iii) their public availability, which supports reproducibility.

For each project, pipeline definitions were collected across multiple commits spanning several months of development activity. Each definition was parsed, instantiated as a metamodel instance, and transformed into a BPMN 2.0 process model. Execution data (job status, duration, failure reasons) were additionally retrieved to evaluate the integration of structural and behavioral information.

---

[3]https://gitlab.com/wireshark/wireshark
[4]https://gitlab.com/veloren/veloren
[5]https://gitlab.com/inkscape/inkscape

**Table 2: Cross-version comparison for Inkscape pipeline (DT output).**

| Metric | $V_1$ | $V_2$ | Δ |
|---|---|---|---|
| Jobs count | 15 | 17 | +2 |
| Pipeline runs | 16 | 100 | — |
| Success rate (%) | 31.2 | 61.0 | +29.8 pp |
| Avg. duration (s) | 2 550 | 2 709 | +6.2% |
| Median duration (s) | 2 795 | 2 766 | −1.0% |
| Build stage avg. (s) | 614 | 453 | −26.2% |
| Avg. queue time (s) | 3.1 | 50.2 | +1 519% |

## 5.2 Model Expressiveness and structural coverage

Across the analyzed projects, the metamodel was able to consistently represent the main structural elements of CI/CD pipelines, including stages, jobs, dependencies, triggers, and execution-related concepts. In particular, both stage-based execution semantics and DAG-based dependencies were successfully captured within a unified representation. This confirms that the metamodel provides sufficient expressiveness to model real-world pipelines with varying levels of complexity. Furthermore, the transformation to BPMN preserved the essential control-flow semantics, making explicit aspects such as parallelism and synchronization that remain implicit in YAML configurations.

## 5.3 Cross-Version Reasoning and Behavioral Traceability

One of the key objectives of the proposed approach is to support reasoning about pipeline evolution. The analysed projects show that representing pipelines as model instances enables structured comparison across versions. These results confirm that the metamodel provides a stable modeling foundation for capturing and reasoning about pipeline evolution.

To illustrate this capability, we present a comparison of two pipeline versions from Inkscape project. The project's Git history contains over 200 commits modifying its CI/CD configuration, spanning from 2017 to 2026. Through the DT, we acquired and analyzed 41 distinct pipeline definition versions covering a 15-month window (October 2024 – January 2026). Among these, we selected two versions that reflect structural refactoring: version $V_1$ (15 jobs, August 2025) and version $V_2$ (17 jobs, September 2025).

**Structural comparison.** The comparison service identified two added jobs (`deps:macos`, `inkscape:android`), one modified job (`inkscape:macos`, across eight fields), and one new template (`.macos`). The refactoring decomposed the monolithic macOS build into separate dependency and application stages linked by an explicit DAG dependency.

**Behavioral impact.** The metrics service aggregated execution data for each version (16 runs for $V_1$, 100 for $V_2$); Table 2 summarizes the results. The build stage average decreased by 26%, consistent with the caching strategy in $V_2$, and the success rate nearly doubled. Failure analysis showed $V_1$ failures split equally between infrastructure and script errors, whereas $V_2$ failures were predominantly script-level (19/21), suggesting improved infrastructure stability.

This example demonstrates how the DT enables a cross-version causal reasoning that would be difficult to achieve manually: structural changes are precisely identified at the model level, and their behavioral consequences are quantified through execution metrics associated with each version. The continuous synchronization mechanism described in Section 4 ensures that this coupling is maintained over time, keeping structural and behavioral views aligned with the evolving state of the pipeline.

## 5.4 Limitations

The proposed metamodel captures the core structural and execution concepts of CI/CD pipelines in the context of GitLab CI. Nevertheless, it does not yet represent composing and reusing pipeline definitions. Since this aspect mainly concern how pipelines are defined and organized, rather than their underlying control-flow semantics, supporting them would require extending the acquisition and parsing layer.

## 6 Related Work

The proposed approach lies at the intersection of two complementary research directions: the use of DTs for software and DevOps processes, and model-driven approaches for representing, analyzing, and evolving CI/CD pipelines.

### 6.1 DT for software and DevOps processes:

Beyond their application to physical systems, DTs have increasingly been explored for modeling, monitoring, and optimizing processes.

In this perspective, [30] introduces a broader vision of DTs that goes beyond the notion of a passive virtual replica toward what they describe as *self-aware* DTs. In their approach, the DT does not merely mimic process behavior, but reasons about it using data-driven models learned from past executions, with the aim of supporting continuous process improvement. To achieve this, they propose a data-driven modeling approach for process optimization based on big-data analytics over historical process data, with the aim of characterizing normal behavior, detecting deviations in real time, and supporting root-cause analysis in line with continuous improvement principles such as Lean and Six Sigma. This vision is illustrated through an industrial manufacturing use case involving a 3D laser cutting process, where the DT is used to detect deviations, support root-cause analysis, and improve process control in a highly variable and data-intensive environment.

Extending the application of DTs to software-intensive systems, [7] explore their use to support the development and automated testing of complex embedded software systems. Their work proposes a formalization of the DT concept using Object-Z and introduces a DT prototype approach that enables engineers to test software systems in a virtual context without requiring a direct connection to the physical system. They further show how such DT prototypes can be integrated into CI/CD pipelines to support automated integration testing and more agile verification and validation practices. Although their approach is applied to software systems, it remains grounded in a DT perspective that still considers the relation to a corresponding physical system.

Similarly, [28] investigate the role of DT in the context of next-generation Software Quality Assurance (SQA). Their work positions

DTs as a key enabler for real-time simulation and validation of software systems, complementing predictive analytics for proactive defect identification and Agile practices for continuous integration into development workflows. Within this framework, the DT acts as a simulation engine that supports ongoing validation of system behavior, allowing quality assurance to shift from a reactive to a more proactive and continuous process.

Overall, while these works demonstrate the potential of DTs in software engineering contexts, they primarily leverage DevOps practices and infrastructures to support the development, testing, and validation of software systems. In contrast, the majority do not consider DevOps processes themselves as the subject of the DT.

A notable exception is [24], which defines DT for software engineering processes as comprehensive, automated co-pilots across the entire DevOps lifecycle. By sensing process data, reasoning through internal models, and actively suggesting improvements, these DTs aim to mitigate the shortage of skilled engineers by automating repetitive tasks. Ultimately, this holistic approach seeks to optimize development workflows, allowing experts to focus on complex challenges while improving overall software quality. However, this contribution remains at a high level of abstraction and does not detail how such DTs can be operationalized through continuous, model-driven representations and service-based analysis of evolving DevOps processes, such as CI/CD pipelines.

Finally, [3] proposes the CI Build process DT (CBDT) framework, which applies DT technology to holistically optimize CI build processes. Arguing that prior research addressed CI challenges, such as long build durations, frequent failures, and flaky builds, in isolation despite their inherent interdependencies, the authors introduce a framework that provides real-time monitoring, ML-based predictive modeling, and what-if scenario analysis across multiple performance metrics simultaneously. The CBDT architecture integrates monitoring capabilities with prescriptive services, enabling automated feedback and continuous improvement of the actual CI build process. The authors also release a replication package including source code and a 25 GB dataset of build jobs from the ten most popular GitLab projects, facilitating further research.

## 6.2 Model-driven approaches for DevOps:

Complementary to these DT-oriented perspectives, another line of work has focused on model-driven approaches for representing, structuring, and analyzing DevOps processes and CI/CD pipelines.

[15] introduces DevOpsML, a conceptual framework for modeling and combining DevOps processes and platforms using MDE principles. Tools, interfaces, and capabilities serve as building blocks for DevOps platform configurations, which are mapped to software engineering processes of arbitrary complexity, with the aim of making DevOps practices accessible to non-technical users.

[11] identifies requirements for a modeling framework for DevOps derived from industry use cases, establishing the need for model-driven representations that capture DevOps workflow structure and evolution.

[27] proposes a model-driven approach based on justification diagrams to assess whether a CI pipeline is coherent with the requirements of its project, elevating CI configurations to a model level for quality evaluation. [10] introduces a multi-platform specification language that abstracts DevOps pipelines from GitHub, GitLab, Bitbucket, Azure DevOps, and Jenkins into a unified representation, accompanied by a curated dataset for cross-platform analysis. Most recently, [16] proposes a metamodel for reengineering CI/CD pipelines following the horseshoe model: vendor-specific pipeline definitions are reverse-engineered into a tool-agnostic intermediate model, from which semantically equivalent pipelines can be regenerated for a different CI/CD platform.

Taken together, these works highlight the relevance of model-driven approaches for structuring and analyzing DevOps pipelines. However, they do not address their integration within a digital twin system.

## 7 Conclusion

In this paper, we presented a model-driven DT for the systematic improvement of DevOps pipelines, using MDE principles to provide structured, process-level representations that support DevOps pipeline analysis and reasoning.

The proposed approach contributes: (i) a pipeline metamodel that provides a unified representation of both the structural definition and runtime execution of CI/CD pipelines, that supports cross-version comparison; (ii) a model transformation that maps CI/CD configurations to BPMN 2.0 process models, making structural properties within a standardized and platform-independent notation; and (iii) a DevOps pipeline DT that operationalizes these modeling foundations within an event-driven system, integrating continuous data acquisition, persistent model management, and aspect-oriented service organization for visualization, structural comparison, and analytical capabilities.

Evaluation on open-source projects shows that the approach captures pipeline complexity and enables meaningful analysis of pipeline evolution and runtime behavior. A next step consists in deploying the DT in an industrial projects to conduct industrial validation.

Future work will also pursue complementary extensions of this foundation. A first direction consists in leveraging the BPMN models to support process simulation and what-if analysis, thereby enabling the evaluation of alternative pipeline configurations and opening the way toward prescriptive DevOps practices. A second direction consists in integrating learning-based services for more intelligent pipeline analysis, including flaky job detection, failure categorization, and failure prediction, in order to support proactive pipeline management. A third direction concerns the generalization of the approach to other CI/CD platforms, such as GitHub Actions, Jenkins, and CircleCI, by adapting the acquisition layer while preserving the core metamodel and transformation logic. Finally, the metamodel itself will be extended to capture additional DevOps process dimensions, such as issue tracking, Kanban boards, and peer review processes, enabling a more comprehensive and unified model-driven DT of the software delivery lifecycle.

## 8 Acknowledgments

We acknowledge the support of the Natural Sciences and Engineering Research Council of Canada (NSERC) [projects RGPIN-2025-06421 and ALLRP 576653] and Mitacs [IT32248].